\documentclass[10pt]{article}
\usepackage[dviwin]{graphicx}

\begin{document}

\begin{center}
{\large NUMERICAL \ SIMULATION \ OF \ STOCHASTIC \\ VORTEX \ TANGLES}

\bigskip

{ \large L.P. Kondaurova, S.K. Nemirovskii, M.V. Nedoboiko \\
\bigskip
Kutateladze Institute of Thermophysics SB RAS,\\
630090 Novosibirsk, Russia}
\bigskip
\bigskip

{\bf Abstract}

\end{center}

\bigskip

We present the results of simulation of the chaotic
dynamics of quantized vortices in the bulk of superfluid He II.
 Evolution of vortex
lines is calculated on the base of the Biot-Savart law. The
dissipative effects appeared from the interaction with the normal
component, or/and from  relaxation of the order parameter are
taken into account. Chaotic
dynamics appears in the system via a random forcing, e.i. we use
the Langevin approach to the problem. In the present paper we
require the correlator of the random force to satisfy the
fluctuation-disspation relation, which implies that thermodynamic
equilibrium should be reached.
In the paper we describe the numerical methods for integration
 of stochastic
differential equation (including a new algorithm for reconnection
processes), and we present the results of calculation of some
characteristics of a vortex tangle such as the total length,
distribution of loops in the space of their length, and the energy
spectrum.

PACS: {\bf 47.32.Cc, 67.40.Vs, 05.40.-a}

%\newpage

\section{Introduction}

Quantized vortices appearing in quantum fluids influence many
properties of the systems. In general, the set of these vortices
represents a chaotic vortex tangle (VT) consisting of separated
vortex loops (closed vortex lines). To describe the influence of
VT, it is necessary to know the statistical description of VT at
various moments of time evolution because at different processes
quantized vortices reveal the different properties:
thermodynamical equilibrium and nonequilibrium (turbulent). For
example quasi-equilibrium features are essential in the process of
fast quenching. In turn, in experiments with thermal flows and/or
counterflows in He II a set of the vortices shows very
nonequilibrium (turbulent type) properties. The statistical
descriptions of these two cases are strongly different. Thus the
distribution of vortex loops on their length $n(l)= {dN(l)}/{dl}$
is governed by the formula [1] $n(l)\sim l^{-5/2}$ in the case
of the thermal equilibrium, while in turbulent helium we have
[2] $n(l)\sim l^{-4/3}$ . There exist many works devoted to
numerical investigations of vortex tangles and to turbulent state
of helium, e.g.[3-10]. Let's note that the mentioned
calculations have been done in the local approximation framework.
Originally simple vortex structures (VS) with time turn into a
very strongly tangled system. if a self-crossing of the filaments
happens in this system, the reconnection
 of vortex line occurs and thus the vortex loops divide or confluence. Reconnections change
the topology of vortex structures and affect the evolution of a
VT. In the papers mentioned above, reconnection was simulated from
the condition of the equality between the local and nonlocal
contributions to the velocity of a vortex filament point, i.e.,
$v_{nl}\approx  \kappa /{2\pi \Delta }\approx v_l\approx {c\kappa
}{\ln \left( R/a_0\right) }/{4\pi R }$, where $\Delta ={2R}/[{c\ln
\left( R/a_0\right) }]$ is the minimal distance between the pair
of vortices, $\kappa $ is the quantum of circulation, $R$ is the
radius of curvature at the given point, $c$ is the constant
($\simeq 1 $), and $a_0$ is the cutting parameter concerning with
the radius of the vortex core. Thus, in those works only the
distance between the points of a vortex line was chosen as the
criterion for reconnection. In our opinion, this is slightly
incorrect approach, because the elements of filaments can go away
from each other and the crossing may not occur. In contrast to the
mentioned above papers, we consider the entire Biot-Savart
equation. Moreover, we take into account possible random
disturbances in the system of vortices. The disturbances are
simulated by addition of a new term to the Biot-Savart equation.
The details see below. This statement is conventional for the
description of dynamical systems with stochastical perturbations.
Moreover, in this work the following condition of reconnections is
proposed: if the elements of a VT have intersected during the
temporal step of the calculation, the reconnection occurs.

\section{ The problem statement and the dynamical equation}

We consider the dynamics of vortex loops in three-dimensional
infinite space. The induced velocity of helium at a point\textbf{\
}$r$ is defined by the Biot-Savart law:
\[
\mathbf{v}(\mathbf{r})=\frac \kappa {4\pi }\int \frac{\left( \mathbf{S}-
\mathbf{r}\right) \times d\mathbf{s}}{\left| \mathbf{S}-\mathbf{r}\right| ^3}
\]
The formulae for the velocity of vortex line points, while without
dissipation, takes the form [3]:
\begin{equation}
\frac{d\mathbf{s}}{dt}=\stackrel{\cdot }{\mathbf{S}_0}=\frac \kappa {4\pi
}\int \frac{\left( \mathbf{S}_1-\mathbf{S}\right) \times d\mathbf{S}_1}{
\left| \mathbf{S}_1-\mathbf{S}\right| ^3}+\frac \kappa {4\pi }\ \ln\left ( \frac
 {2\sqrt{S_{+}S_{-}}}{e^{1/4}\cdot a_0}\right)\ \mathbf{S}^{\prime }\times \mathbf{%
S}^{\prime \mathbf{\prime }},
\end{equation}
where $\mathbf{S}(\xi ,t)$ is the radius-vector of the vortex line points, $\xi $ is
a parameter, in this case the arclength, $\mathbf{S}^{\prime }$ is the arclength derivative,
$S_{+}$ and $S_{-}$ are the lengths of two adjacent line
elements that hold the point $\mathbf{s}$ between, and the prime denotes
differentiation with respect to the arclength $\xi .$ The second term of Eq.(1) is the local part of the velocity the first term is the nonlocal part
obtained by integration on the rest of the vortex line and on all of the
other loops. Taking into account the frictional force of vortices and the
normal component of helium and the rapidly fluctuating random term
(Langevin's force), we obtain the equation for the dynamics of a point of the
vortex line:
\begin{equation}
\stackrel{\cdot }{\mathbf{S}}=\stackrel{\cdot }{\mathbf{S}_0}+\alpha \left(
\mathbf{S}^{\prime }\times \left( \mathbf{v}_n-\stackrel{\cdot }{\mathbf{S}}
_0\right) \right) -\alpha ^{\prime }\mathbf{S}^{\prime }\times \left[
\mathbf{S}^{\prime }\times \left( \mathbf{v}_n-\stackrel{\cdot }{\mathbf{S}}
_0\right) \right] +A(t,\xi )
\end{equation}
where $ \quad \left\langle A(t,\xi )\right\rangle =0, \quad
\left\langle A_i(t_1,\xi _1)A_j(t_2,\xi _2)\right\rangle \ =D\
\delta _{ij}\delta \ (t_1-t_2)\delta \ (\xi _1-\xi _2)=D\ \delta
_{ij}\delta \ \ (\xi _1-\xi _2)\ \left\langle n(t)\ n(t^{\prime
})\right\rangle , $ $i,j$~are the spatial components; $t_1,t_2$
are the arbitrary time moments; $\xi _1,\xi _2$  define any points
on the vortex line; $D$ is the intensity of the Langevin's force;
$\alpha ,\alpha ^{\prime }$  are the friction coefficients; $n(t)$
is the Gaussian white noise with $\left\langle n(t)\ \right\rangle
=0,$ $\left\langle n(t)\ n(t^{\prime })\right\rangle =\delta \
(t-t^{\prime }).$ Let's assume  further that the difference
between the normal and the superfluid velocities of helium equals
to zero $\mathbf{v}_n=0$ and neglect the term with $\alpha
^{\prime }$. This statement correspond to absence of a heat flow.
Thus, finally we obtain the dynamical equation for a vortex line:
\begin{equation}
\stackrel{\cdot }{\mathbf{S}}=\stackrel{\cdot }{\mathbf{S}}_0-\alpha
\mathbf{S}^{\prime }\times \stackrel{\cdot }{\mathbf{S}}_0
+\sigma (\xi)n(t),
\end{equation}
In the integral form Eq.(3) is as follows:
\begin{equation}
\mathbf{S}(t,\xi ) =
=\mathbf{S}(t_0,\xi )+\int\limits_{t_0}^{t}B(\xi ,t)dt+\sigma (\xi)
\int\limits_{t_0}^{t}dW,
\end{equation}
where $B(\xi ,t)=\stackrel{\cdot }{\mathbf{S}}_0-\alpha \mathbf{S}^{\prime
}\times \stackrel{\cdot }{\mathbf{S}}_0$, and $W(t)=\int\limits_{t_0}^{t
}n(t^{\prime })dt$ is the standard Wiener process.
 In our model the initial conditions is six completely symmetrical
rings with orientation making the total impulse of the system equal to zero.

\section{The numerical algorithm and the description of reconnections}

The Eq.(4) was solved by the Euler method:
\[
S_{n+1}=S_n+hB(\xi ,t_n)+\sqrt{h}\sigma (\xi )\eta _n,
\]
where $S_n$ is the approximate solution of the equation in a mesh point on
time $t_{n,}$ $h$ is the integration step on time in a mesh point $t_n,$
 $\left\{ \eta _n\right\}$ is the consistency of independent between
itselves normal random vectors with independent between itselves
components in the aggregate $\eta _{n,j}$ $(j=1,2,3),$ having zero
expectation value and the dispersion is 1. The component of vector
$\eta _n$ was calculated by formula: $\eta _{n,j}=\sqrt{-2\ln
\alpha _1}\cos (2\pi \alpha _2)$, where $\alpha _1$ and $\alpha
_2$ is random numbers from the interval (0,1) obtained by a
pseudorandom-number generator. The Euler method is the first order
on the mean-square approximation in the time step. The functions
$\mathbf{S}^{\prime },\mathbf{S}^{\prime \mathbf{\prime }},
\stackrel{\cdot }{\mathbf{S}}_0$ were calculated as in paper [3].
To keep the calculation procedure coherent, points on the vortex
line were added and removed was carry out as in the work [10].

The first step in the modeling of a reconnection process is selection of
point pairs that are the candidates for reconnection.
 After the pairs was defined, it was
assumed that the line segments between each of the pairs were moving with a
constant velocity (according $V_i,V_j$) during the time step, as illustrated
in Fig. 1.

\begin{figure}[htbp]
\unitlength=1mm
\begin{picture}(80,40)
%{reconec0.gif}
\put(45,37)%{reconec0.}
{\special{em:graph 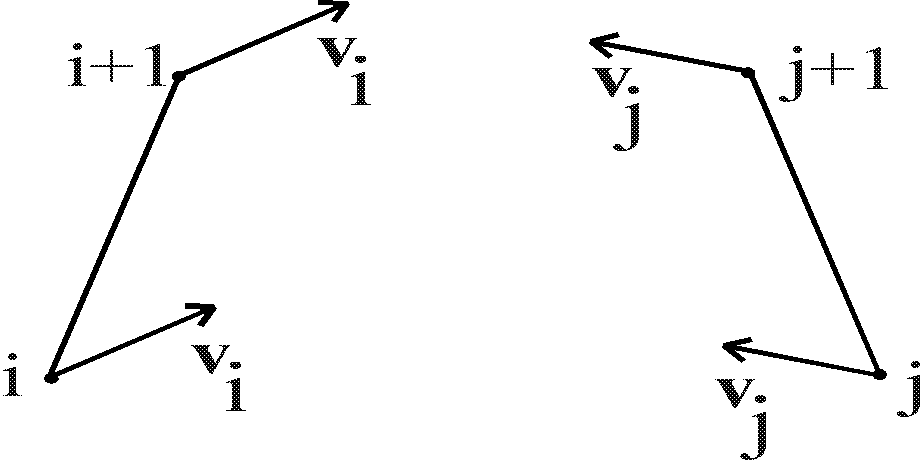}}
%{em:graph reconec0.gif}}
\end{picture}
%\centering
%\includegraphics[type=bmp]{reconec0.gif}
\caption{The elements of vortex lines that must reconnect}
\end{figure}

From the compatibility of equations
\[
x_i+V_{x,i}h+(x_{i+1}-x_i)s1=x_j+V_{x,j}h+(x_{j+1}-x_j)s
\]
\[
y_i+V_{y,i}h+(y_{i+1}-y_i)s1=y_j+V_{y,j}h+(y_{j+1}-y_j)s
\]
\[
z_i+V_{z,i}h+(z_{i+1}-z_i)s1=z_j+V_{z,j}h+(z_{j+1}-z_j)s
\]
\[
0\leq s1\leq 1;0\leq s\leq 1
\]
it was determine the meeting of these line segments during the time step. Here
$(x_i,y_i,z_i,$ $x_{i+1},y_{i+1},z_{i+1});$ $(x_j,y_j,z_j,$
$x_{j+1},y_{j+1},z_{j+1})$ are the coordinates of the first and the second pairs of
points, accordingly; $V_{x,i},V_{y,i},V_{z,i};V_{x,j},V_{y,j},V_{z,j}$ are the
projections of the velocities of the points and the line segments on the
coordinate axis. If the line segments had met, the reconnection
occurs. Thus, if originally the points belong to the same loop, a pair of
new loops was generated. Otherwise the confluence of the loops occurs.

\section {The results}

The initial radii of rings were $R=2\cdot 10^{-5}m$ . The initial condition
was chosen in a way that the total impulse of the system was equal to
zero. The rings were situated symmetrically at equal distance in pairs around
 the coordinate origin. The distance between them was $d=10^{-5}m.$ The parameters in
Eq.2 are $\alpha =0.0098,$ $D=4\cdot 10^{-5}m/s.$

Simulation was performed with a constant temporal step $h=5\cdot
10^{-8}$s and initial steps along the vortex line was $\Delta l_0=2\pi \cdot
10^{-7}$ m. The steps $\Delta l$ along the vortex line was controlled later by
the procedure of inserting and removing of points, so that ${\Delta l_0}/
2$ $\leq \Delta l\leq 2\Delta l_0$.  It follows from Eq.$4$ that small
(or those with a high curvature) loops move very rapidly and their
dissipation (decrease in the size) is very high due to the friction.
Therefore, small loops were removed during our calculations. Kinks
appearing on vortex lines were removed also. For our case, the
loops were cancelled if there were less than 5 points.

Vortex configurations at various time moments are presented in
Fig.2.
\begin{figure}[htbp]
\unitlength=1mm
\begin{picture}(80,130)
\put(-5,125){\special{em:graph 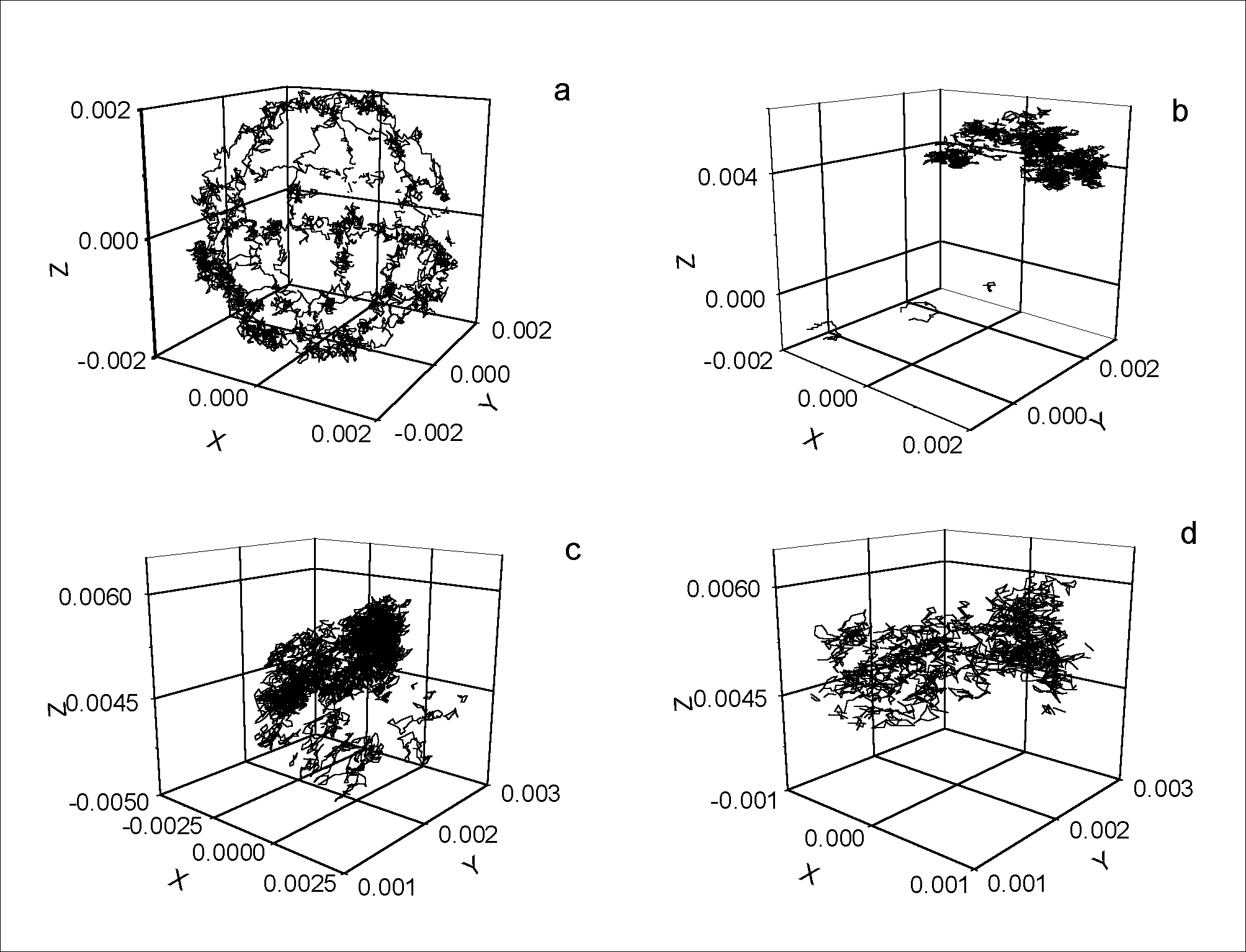 }}
\end{picture}
\bigskip
\caption{
 The vortices at different times: a)  $t=4.30\cdot 10^{-5}s$,
b) $ t=9.792\cdot 10^{-4}s$, c) $t=1.0892\cdot 10^{-3}s$,
 d) $t=1.1444\cdot 10^{-3}s.$ The configurations are plotted
in a three-dimensional view.}
\end{figure}

 One can see a vortex structure with drastic evolution in
time. After numerous time steps the system evolved into separated
vortex tangles. It was noted that during the evolution, the vortex
tangles arose and then vanished in different places. It resembles
 us the favor intermittency phenomenon
in classical turbulence.

From the simulation, several quantities were calculated and
plotted as functions of  time (see Fig.3). One can see the phases
of evolution for vortex structures. At the beginning, the total
length, averaged curvature, and the vortex lines density increase
at a steady volume. After a certain value of the density has been
archived, many of small loops developers and the VT begin to
decay. Later, one can observe a tendency to a fluctuating steady
state. However, after the time $t= 9.792\cdot 10^{-4}s$, the
volume suddenly begin to decrease. It is conform to disappearance
of detached loops (see Fig.2 c, d ). Then again the fluctuating
steady state is reached.
\begin{figure}[htbp]
\unitlength=1mm
\begin{picture}(180,45)
\put(3,40){\special{em:graph 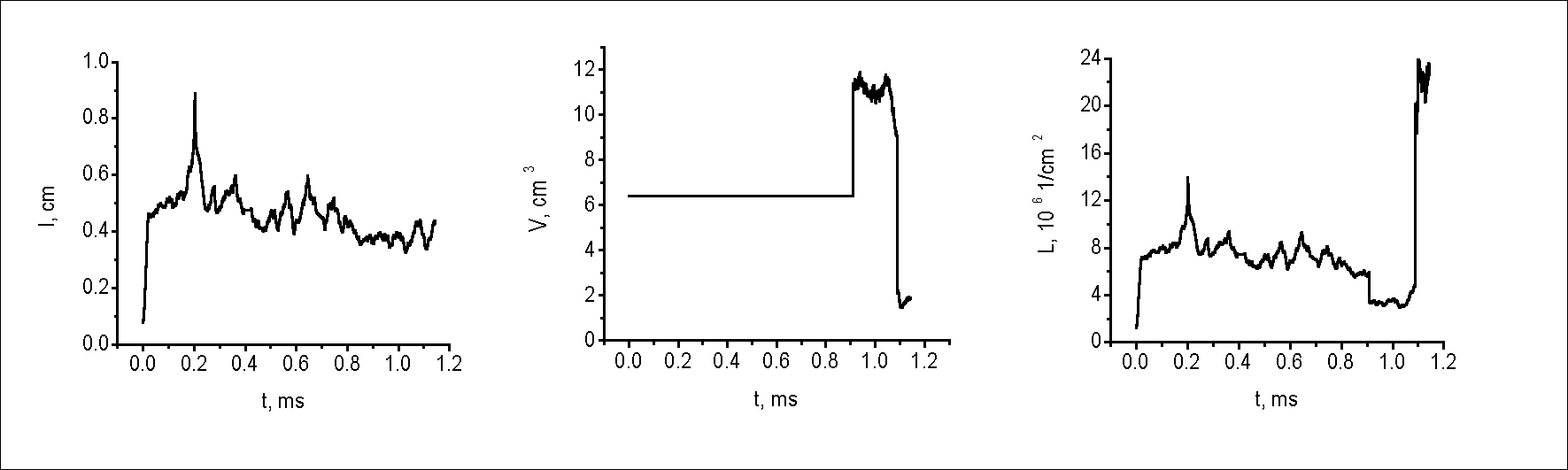}}
\end{picture}
\caption{
 Plots of the total lines l, the volume V, filed with
vortices, and the vortex lines density L as functions of time.}
\end{figure}

We made also calculations for some statistical characteristics of
a VS. The distribution of the loop number vs. its length and the
VC energy spectrum were calculated. Figure 4 shows the
distribution of vortex loops on their length at the time region
$t\approx 1.1443\cdot 10^{-3}s.$ Decreasing of the loops amount is
described by the following function: $n(l)dl\sim l^{-0.93}$ . The
same dependance was observed both for different times and for each
vortex tangle within the vortex structure. It is difficult to
extract physically meaningful results from this data. On the one
hand, the equilibrium has been reached. On the other hand, the
distribution of vortex loops disagrees with the results obtained
for thermal equilibrium and turbulent stationary state.
\begin{figure}[htbp]
\unitlength=1mm
\begin{picture}(80,65)
\put(20,60){\special{em:graph 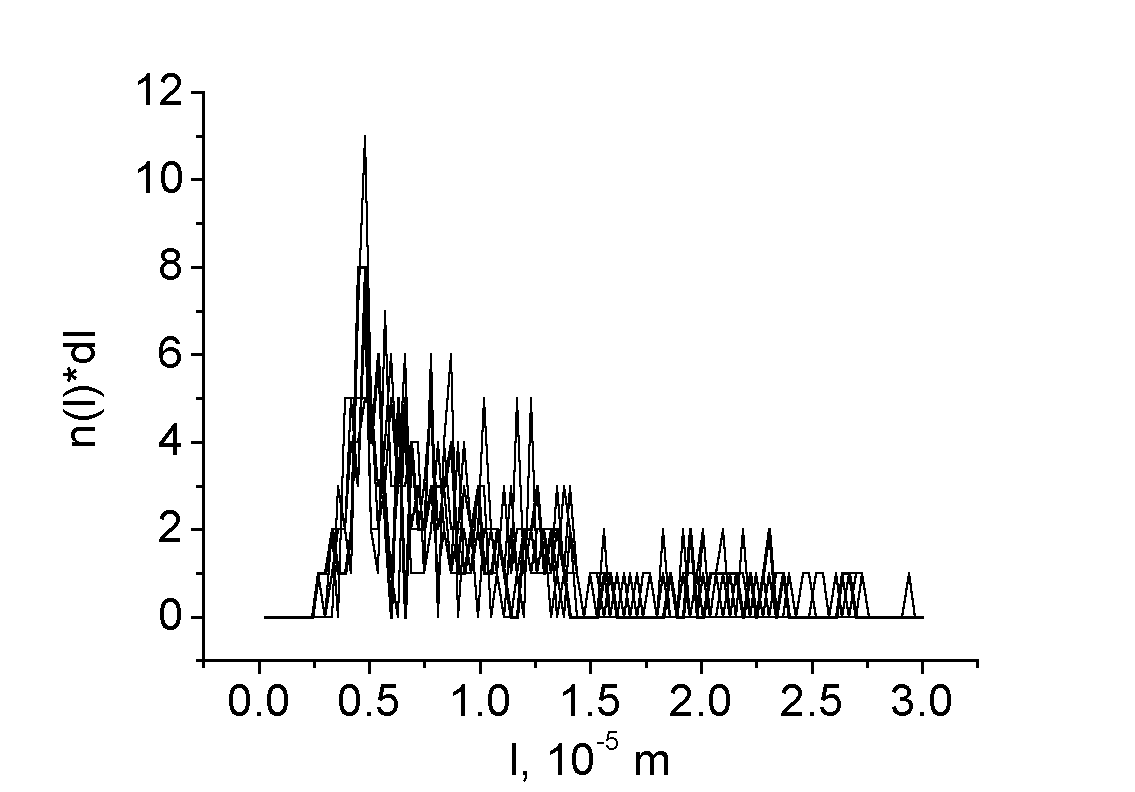}}
\end{picture}
\caption{
 The distribution of the vortex loops on their length
at the time region
$t\approx 1.1443\cdot 10^{-3}s. $}
\end{figure}

The average kinetic energy of flow induced by a vortex loop can be evaluated
as follows [11]:
$$E=\left\langle\int \frac{\rho _sV_s^2}{2}d^3r\right\rangle =
\int\limits_{k}\frac{d^3k}{4\pi k^2}\left\langle 2\pi \rho _s\kappa ^2
\int \limits_0^{L^{\prime }}\int \limits_0^L
{\mathbf{S}}(\xi _1)\stackrel{\cdot }{\mathbf{S}}(\xi
_2)e^{-ik(\mathbf{S}(\xi _1)-\mathbf{S}(\xi _2)}d\xi _1d\xi _2\right\rangle =
\int \limits_0^{\infty }E(k)dk $$
$$E(\mathbf{k})=\frac{\rho _s\kappa ^2}{2(2\pi )^3}\int \frac{d\Omega _k}{k^2}%
\int\limits_0^L
\int\limits_0^{L^{\prime }}\stackrel{\cdot }{\mathbf{S}}(\xi _1)\stackrel{\cdot }{\mathbf{S}}%
(\xi _2)e^{-i\mathbf{k}(\mathbf{S}(\xi _1)-\mathbf{S}(\xi _2)}d\xi _1d\xi _2,
$$
where $d\Omega _k=k^2\sin \Theta _kd\Theta _kd\Phi _k$ is the elementary volume,
$\mathbf{k}$ is the wave vector, and $\rho _s$ is the superfluid density.
For the isotropic case, the spectral density is expressed as follows:
$$
E(k)=\frac{\rho _s\kappa ^2}{(2\pi )^2}\int
\limits_0^L
\int\limits_0^{L^{\prime }} \frac{\sin (k\left| \mathbf{S
}(\xi _1)-\mathbf{S}(\xi _2)\right| )}{k\left| \mathbf{S}(\xi _1)-\mathbf{S}
(\xi _2)\right| }\left( d\mathbf{S}(\xi _1)\cdot d\mathbf{S}(\xi _2)\right) ,
$$
where $k$ is the wave number. It is seen that there exist different regions of
wave number $k$: small with respect to $(V)^{1/3}$ , high with
respect to $1/\sqrt{L}$, and intermediate values. In the region of
small wave numbers, $E(k)\propto k^2,$ and for high numbers, $E(k)\propto
k^{-1}.$

Figure 5 shows  our numerical results of the VC spectral density for the time $t=1.14445\cdot 10^{-3}s$ .
In the region of small wave numbers, simulation data
fits the approximation $E(k)\propto k^2$. In the region of high wave numbers,
it is difficult to extract a regularity from this data, since the spectral density
and the numerical error are of the same scale.
In the intermediate region, the energy decreases according to the formula: $%
E(k)\propto k^{-2.9}$.
\begin{figure}[htbp]
\unitlength=1mm
\begin{picture}(180,65)
\put(20,60){\special{em:graph 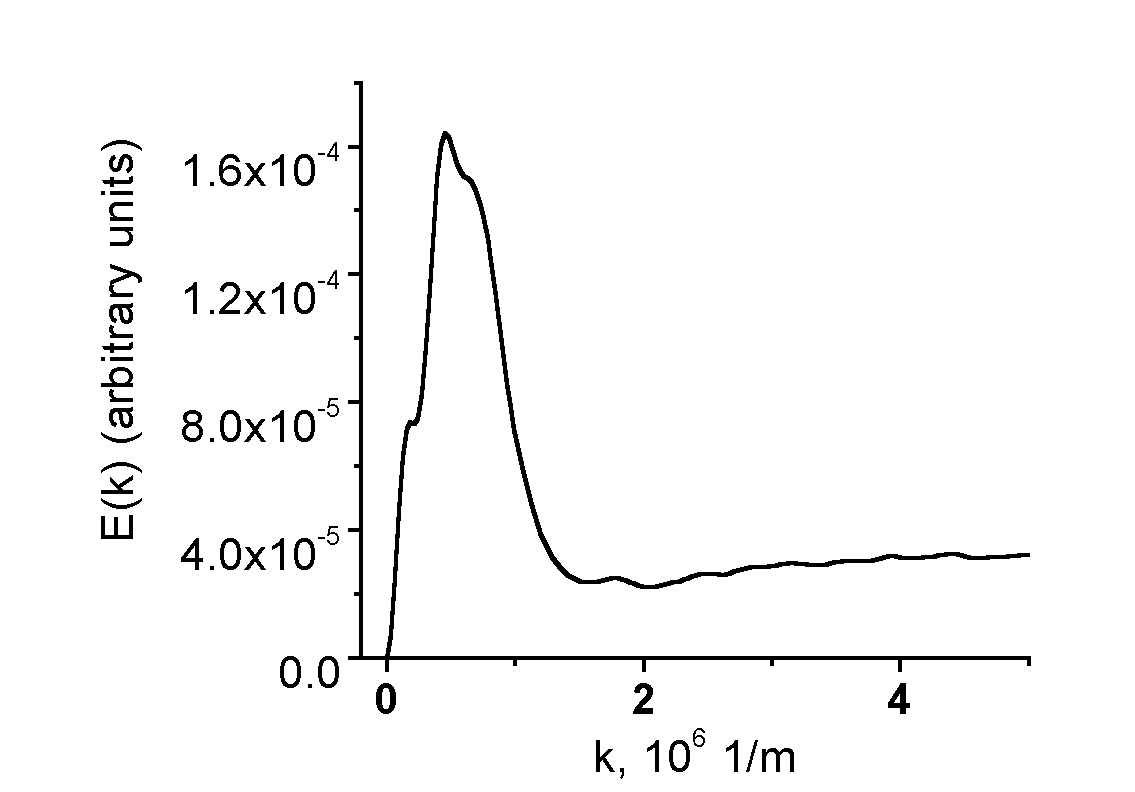}}
\end{picture}
\caption{ The energy spectrum of the vortex tangle at
$t= 1.1445\cdot 10^{-3}s. $}
\end{figure}

\section{CONCLUSION}
The obtained numerical results demonstrate that initially smooth vortex
rings transform into a highly chaotic vortex tangle. In spite of that total
length fluctuates about a constant value, we think that the  thermal equilibrium
state has not been reached yet. For instance, the  spectral density of the
energy agrees with the equipartition law only for a small k-zone. The
vortex loop distribution on their length also differs from the
one expected for a thermal equilibrium. This state is closer to the turbulent case
observed by other authors. During evolution the vortex tangles arise and
disappear in different places. It resembles the favor intermittency phenomenon
in classical turbulence. Our preliminary simulation demonstrates that
 the Langevin approach applied in this paper is a very promising method for
study of chaotic vortex structures.

 The work is supported by INTAS grant 2001-0618.

\end{document}